# Tunable dual-comb from an all-polarization-maintaining single-cavity dual-color Yb:fiber laser

JAKOB FELLINGER[1]*, ALINE S. MAYER[1], GEORG WINKLER[1], WILFRID GROSINGER[1], GAR-WING TRUONG[2], STEFAN DROSTE[3], CHEN LI[3], CHRISTOPH M. HEYL[3,4], INGMAR HARTL[3] AND OLIVER H. HECKL[1]*

[1]*Christian Doppler Laboratory for Mid-IR Spectroscopy and Semiconductor Optics, Faculty Center for Nano Structure Research, Faculty of Physics, University of Vienna, Vienna, Austria*
[2]*Crystalline Mirror Solutions LLC, Santa Barbara, California 93101, USA*
[3]*Deutsches Elektronen-Synchrotron DESY, Notkestraße 85, 22607 Hamburg, Germany*
[4]*Helmholtz-Institute Jena, Fröbelstieg 3, 07743 Jena, Germany*
*jakob.fellinger@univie.ac.at*
*oliver.heckl@univie.ac.at*

**Abstract:** We demonstrate dual-comb generation from an all-polarization-maintaining dual-color ytterbium (Yb) fiber laser. Two pulse trains with center wavelengths at 1030 nm and 1060 nm respectively are generated within the same laser cavity with a repetition rate around 77 MHz. Dual-color operation is induced using a tunable mechanical spectral filter, which cuts the gain spectrum into two spectral regions that can be independently mode-locked. Spectral overlap of the two pulse trains is achieved outside the laser cavity by amplifying the 1030-nm pulses and broadening them in a nonlinear fiber. Spatially overlapping the two arms on a simple photodiode then generates a down-converted radio frequency comb. The difference in repetition rates between the two pulse trains and hence the line spacing of the down-converted comb can easily be tuned in this setup. This feature allows for a flexible adjustment of the tradeoff between non-aliasing bandwidth vs. measurement time in spectroscopy applications. Furthermore, we show that by fine-tuning the center-wavelengths of the two pulse trains, we are able to shift the down-converted frequency comb along the radio-frequency axis. The usability of this dual-comb setup is demonstrated by measuring the transmission of two different etalons while the laser is completely free-running.

## 1. Introduction

Dual-comb (DC) spectroscopy [1] has emerged as a versatile technique combining fast data acquisition, high-resolution and broadband spectral coverage [2]. Typically, DC systems involve two separate mode-locked lasers emitting pulse trains with slightly different pulse repetition rates. Those pulse trains correspond to two frequency combs with slightly different comb line spacings. Optical beating between the two combs leads to a down-conversion of the optical frequencies into the radio-frequency range. Since this down-conversion takes place on a simple photodiode sensor, DC spectroscopy does not require complex spectrometer assemblies such as virtually imaged phase arrays [3] or Fourier transform spectrometers [4]. A critical aspect, however, is the mutual coherence between the two optical combs. When the combs originate from two independent lasers, active stabilization can quickly become a complex and difficult task. Different methods have already been presented to reduce the complexity of dual-comb setups, for example by phase-locking the two frequency combs to an external cavity diode laser [5] or using adaptive sampling techniques [6–9]. An additional simplification is the generation of two pulse trains using a single laser cavity: mutual coherence due to common-mode noise cancellation enables spectroscopy with a free-running laser [10]. Single-cavity dual-combs have been obtained using various mechanisms, e.g. by separating the two pulse trains using different travel directions [11–13], polarization [14,15] or branched optical paths in a birefringent crystal [16,17].

Here, we focus on a scheme that consists of using a single-cavity dual-color/dual-comb fiber laser, where a single laser cavity emits two pulse trains with different center wavelengths [18]. Non-zero intra-cavity dispersion leads to a difference in repetition rate for the two emitted pulse trains. Based on state-of-the-art fiber lasers, the dual-color/dual-comb approach [19–28] shows great potential for spectroscopic measurements without the need for active stabilization of the comb parameters [22,23,27]. For instance, Zhao et al. described a carbon nanotube mode-locked dual-color erbium doped fiber laser [20], which then further evolved into a single-cavity dual-comb and enabled free-running spectroscopy [22]. Subsequently, Liao et al. demonstrated a thulium-doped nonlinear amplifying loop mirror (NALM) mode-locked dual-color dual-comb laser, pushing this technique towards the mid-infrared spectral region [23]. Furthermore, Li et al. focused on improving the stability of such systems to enable fieldable spectroscopy. They reported an all polarization-maintaining (PM) dual-wavelength mode-locked erbium fiber laser, using a Sagnac loop filter, reaching a repetition rate around 40 MHz with a difference in repetition rate ($\Delta f_{\text{rep}}$) rate of around 900 Hz, leading to a non-aliasing dual-comb bandwidth of 0.9 THz [25]. More recently, Chen et al. closed the spectral gap between the erbium and the thulium emission spectra using nonlinear broadening in a fiber [27].

In recent work, we have demonstrated the implementation of a mechanical spectral filter to generate a tunable dual-color laser [26]. The method was implemented in a nonlinear polarization evolution (NPE) mode-locked Yb:fiber laser operating at 23 MHz. Our dual-color laser scheme offers two key features: dynamic adjustment of the spectral filter and tuning of the difference in repetition rates.

In this work, we combine these features with the benefits of an all-PM-NALM laser design. Implementing tunable spectral filtering inside a PM-laser instead of an NPE system leads to significantly higher stability and reliability. In addition, the tuning possibilities offer direct control over important dual-comb properties – the non-aliasing bandwidth and the acquisition rate. The latter is given directly by the difference in repetition rates $\Delta f_{\text{rep}}$, while the non-aliasing bandwidth $\Delta \nu$ – i.e. the maximally allowed optical overlap of the two combs before spectral aliasing occurs – must fulfill the following condition [1]:

$$\Delta \nu \leq \frac{f_{\text{rep},1}^2}{2\Delta f_{\text{rep}}}. \tag{1}$$

Furthermore, we increased the repetition rate by a factor of two compared to previously reported PM dual-color lasers system [25], which also helps to achieve a higher non-aliasing bandwidth (see Eq. 1). Since our laser is based on an ytterbium gain fiber, we reach a total output power of 9.2 mW, which is a factor of 5 higher than reported in previous PM dual-color lasers schemes based on erbium gain fibers. The power scaling potential of such Yb-based platforms makes them highly suitable for nonlinear wavelength conversion, which is a promising way to extend the wavelength coverage of dual-comb spectroscopy [27,29].

## 2. Dual-comb setup

The mode-locking mechanism of our PM Yb:laser is based on the method described by Hänsel et al. [30], where a nonlinear amplifying loop mirror [31] is combined with a nonreciprocal phase shifter. NALM mode-locking has the important advantage that it does not rely on nonlinear polarization evolution, hence allowing the use of PM fibers. The PM fiber design decreases the sensitivity to environmental perturbations, leading to reliable, repeatable and self-starting mode-locked operation. The complete dual-comb setup is shown in Fig. 1. The laser consists of a fiber portion and a linear free-space arm. The fiber section includes a PM wavelength division multiplexer (WDM), a 45-cm long PM-Yb:doped fiber (CorActive YB 401-PM) diode-pumped at 976 nm (BL976-PAG900) and a birefringent polarization beam combiner collimator (PBCC). The free-space arm involves a Faraday rotator (FR), two quarter wave plates (QWP), a half-wave plate (HWP), a polarization beam splitter cube (PBS) and a grating compressor (Wasatch Photonics, 800 lines/mm, angle-of-incidence 24.3° @ 1030 nm) for group delay dispersion (GDD) compensation. Spectral filtering is introduced in the grating compressor using a blade-shaped beam block [26].

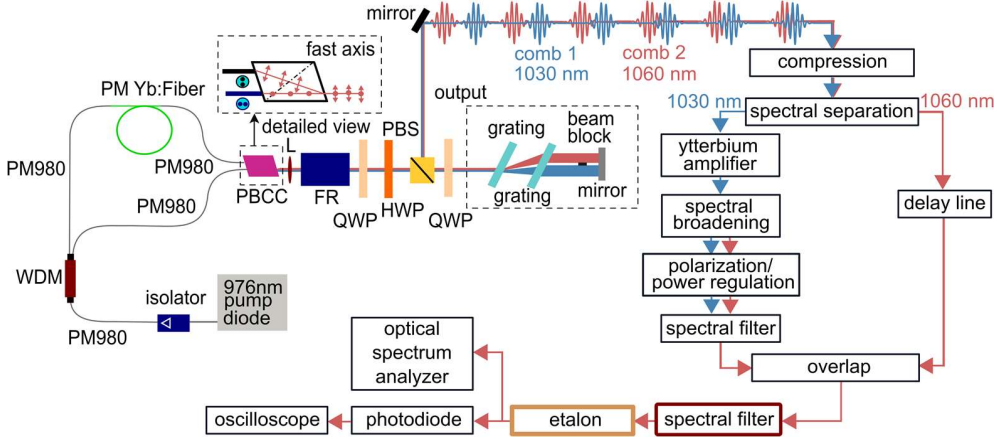

Fig. 1. Overview of the all-PM NALM mode-locked single-cavity dual-color/dual-comb setup. Due to the mechanical spectral filter, the laser oscillator emits two pulse trains with different repetition rates around 77 MHz and center wavelengths around 1030 nm and 1060 nm, respectively. The output of the dual-color laser is spectrally separated using a dichroic filter: The pulse centered around 1030 nm is amplified and nonlinearly broadened. The pulse centered around 1060 nm is delayed using a passive fiber. Subsequently, spatial overlapping in a 50:50 fiber splitter/combiner leads to the generation of a dual-comb interferogram. Bandpass filtering of the light is applied to avoid spectral aliasing. The feasibility of spectral measurement is demonstrated by measuring the transmission of different etalons. The light is detected by a simple photodiode and measured with an oscilloscope.

By blocking the center part of the spatially dispersed light in the grating compressor, we force the laser to emit light in two separate regions of its gain spectrum. The operation point is initially found by rotating the wave-plates and ramping up the pump power until both spectral regions mode-lock individually. Note that once the optimum wave-plate position has been found, the wave-plates can be clamped and no additional wave-plate rotation is required when cycling the laser on and off. Reliable and repeatable mode-locking is then achieved just by ramping up the pump power and fine-tuning the spectral filter, i.e. the position of the beam block.

## 3. Dual-color operation regimes

Our all-PM dual-color laser emits two pulse trains with slightly different repetition rates ($\Delta f_{\text{rep}}$ ~1-10 kHz) around ~77 MHz. Changing the grating separation leads to a change in the intra-cavity group delay and group delay dispersion (GDD), and hence enables us to tune the difference in repetition rates. Figure 2 shows the dual-comb operating at three different grating separations, i.e. with $\Delta f_{\text{rep}}$ = 1.2 kHz, 2.6 kHz and 9 kHz. At $\Delta f_{\text{rep}}$ = 1.2 kHz, the non-aliasing bandwidth amounts to 2.4 THz, corresponding to about 10 nm at a center wavelength of 1060 nm, which is more than 2.5 times higher than reported for previous PM dual-color/dual-comb setups.

We measured the intra-cavity dispersion using a technique described by Knox [32]. Since our laser cavity already contains a built-in spectral manipulator, we used it to shift the center wavelength of the laser (in single pulse/color mode) to different spectral positions within the laser gain bandwidth and recorded the repetition rate of the pulses using a spectrum analyzer with a resolution bandwidth of 1 Hz. We calculated the group delay

$$\frac{\partial \phi}{\partial \omega} = T_g = \frac{1}{f_{rep}},\tag{2}$$

and fitted the data with a second-order polynomial. The first derivative of this fit then yields the corresponding GDD, i.e.

$$\frac{\partial^2 \phi}{\partial \omega^2} = \frac{\partial T_g}{\partial \omega} = \frac{\partial T_g}{\partial \lambda}\frac{\partial \lambda}{\partial \omega} = -\frac{\lambda^2}{2\pi c}\frac{\partial T_g}{\partial \lambda}\tag{3}$$

where $\omega$ denotes the angular frequency, $\lambda$ is the wavelength and $c$ is the speed of light.

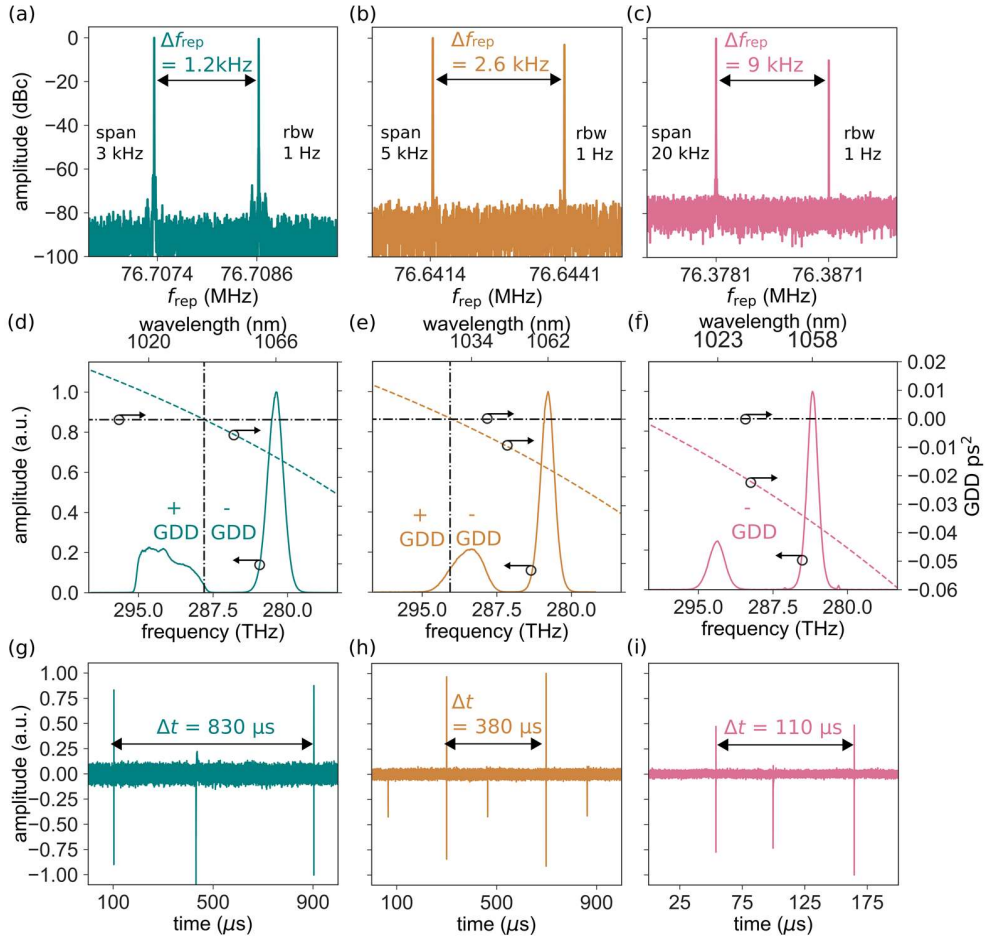

Fig. 2. Dual-comb setup operating with three different grating separations, i.e. different values of $\Delta f_{rep}$, resulting in either higher non-aliasing bandwidth or faster acquisition times. (a-c) Radio frequency trace of the laser outputs recorded with a Keysight PXA N9030B. (d-f) Optical output spectra before spectral separation recorded with an optical spectrum analyzer (ANDO AQ6315A) and measured intra cavity dispersion. The dashed black lines show the zero-GDD crossing points. (g-i) Time trace recorded on an oscilloscope (LeCroy WavePro 760Zi) showing the interferograms that are separated in time by $\Delta t = 1/\Delta f_{rep}$. The spurious signals between the interferograms are the result of intra-cavity pulse collisions and will be discussed in section 4.1.

Figures 2(d)-(f) show the intra-cavity GDD and the dual-color output spectra for the three operation points mentioned above. For the operation point with the largest grating separation (Fig. 2 (f)), both pulses propagate in the negative (i.e. anomalous) dispersion regime and show soliton-like sech$^2$-shaped spectra. However, when reducing the grating distance, the intra-cavity dispersion crosses zero right between the two different spectra of the pulses. Consequently, while the pulse centered around 1060 nm is still running in the negative dispersion regime, the pulse centered around 1030 nm operates in the positive dispersion regime and is spectrally broader.

### 4. Dual-comb characterization

In the following, we will focus on describing the dual-comb operation point at $\Delta f_{rep}$ = 2.6 kHz (Fig. 2(b)), which provides a compromise between non-aliasing bandwidth and acquisition time. The two pulse trains are spectrally located around 1030 nm (comb 1, blue in Fig. 3 and subsequent figures) and 1060 nm (comb 2, red). The two pulses have a full width at half maximum (FWHM) of ~15 nm (pulse @ 1030 nm) and ~7 nm (pulse @ 1060 nm), see Fig. 3(a).

After the cavity, the two pulses are simultaneously compressed by a grating compressor. To avoid any crosstalk between the two pulse trains during amplification and nonlinear broadening, we separate the two pulses after the grating compressor using a short-pass filter with a cut-off wavelength of 1050 nm. The pulse centered around 1030 nm is amplified in a PM single-mode amplifier and spectrally broadened using a nonlinear fiber (NKT-SC-5.0-1040) directly spliced to the amplifier. The nonlinearly broadened spectrum is shown in Fig. 3(a). After spectral broadening, the light is coupled out into a free-space section. Since the nonlinear fiber is not PM, we use a quarter-wave plate, a half-wave plate and a PBS to linearly polarize the light. The half-wave plate in combination with the PBS is also used to attenuate the amplified and spectrally broadened light to optimize the contrast of the beating signal. To filter out the spectral components which are not overlapping with the pulse centered around 1060 nm we use an additional long pass filter with a cut-on wavelength of 1050 nm. The two now spectrally overlapped pulses are then spatially overlapped in a 50:50 fiber coupler. Nonlinear effects in the fiber coupler are not critical, since at this point in the setup we have an average power per pulse train of 1-2 mW and due to dispersion in the previous fiber section (amplifier and delay line), the pulses are not compressed anymore.

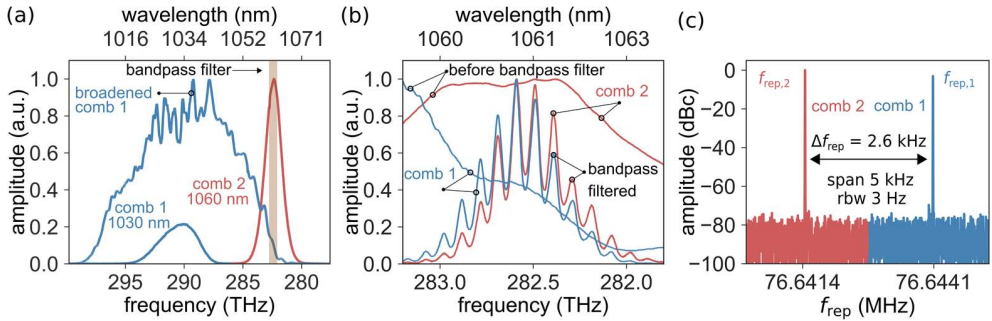

Fig. 3. (a) Output spectrum of the dual-color laser before spectral separation and spectrum of the broadened 1030-nm pulse recorded with an optical spectrum analyzer (ANDO AQ6315A). (b) Spatially overlapped dual-comb output after amplification and spectral broadening, before and after filtering the light using a 3-nm bandpass filter. The modulations on the spectrum are caused by the bandpass filter. (c) Radio frequency spectrum of the laser output. Note that the data shown represents one measurement and that the coloring is merely a guide to the eye.

With a nominal repetition rate $f_{rep,1}$ around 77 MHz, a difference between the two repetition rates of $\Delta f_{rep}$ ~ 2.6 kHz, the non-aliasing dual-comb spectral bandwidth is calculated to be ~1.2 THz, which corresponds to ~ 4 nm around a center wavelength of 1063 nm. To avoid aliasing, we bandpass-filter the light around a center wavelength of 1063 nm with a FWHM of

3 nm. This filter imprints a modulation on the spectra of the two combs, see Fig. 3(c), which however does not impact the spectroscopy experiments as it cancels out when doing a baseline subtraction (see section 7). Finally, we measure the light using a simple photodiode (Thorlabs PDA05CF2), whose signal is then low-pass filtered (48 MHz) to remove the individual comb repetition rates $f_{rep,1}$ and $f_{rep,2}$ before being recorded on an oscilloscope (LeCroy WavePro 760Zi). Figure 2(h) shows an example of an oscilloscope time trace with the interferograms occurring with a periodicity of $T = 1/\Delta f_{rep} = 380$ μs. Note that the spurious signals appearing between the interferograms will be discussed in the next section below.

By performing a fast Fourier-transform on the single interferograms, we can retrieve the down-converted radio-frequency spectra. In order to exploit the full non-aliasing bandwidth in spectroscopy experiments, i.e. to avoid interference with higher-order down-converted frequency combs, the center of this down-converted radio frequency should ideally be located at $f_{rep}/4$. The absolute position of the down-converted comb is related to the difference in carrier-envelope-offset (CEO) frequency of the two combs and is usually not an easily accessible parameter. In this setup however, the mechanical spectral filter can be exploited to tune the down-converted radio frequency comb. The mechanism can be explained as follows: slightly tuning the beam block position results in small changes of the center wavelength of the two pulses, see Fig. 4(a), as well as in their corresponding repetition rates (Fig. 4(b)). The shift is more pronounced for the 1060-nm pulse, which lies further from the cavity zero-dispersion point (see Fig. 2(e)). Although the wavelength shifts are small, they induce a relative change in the dispersion experienced by the two pulses. As a consequence, the difference in CEO changes, which allows for the down-converted radio frequency comb to be shifted considerably along the radio frequency axis (Fig. 4(c)). This useful feature is an additional advantage of mechanical filtering and was exploited to make use of the full theoretical non-aliasing bandwidth in the dual-comb measurements that will be presented in section 5.

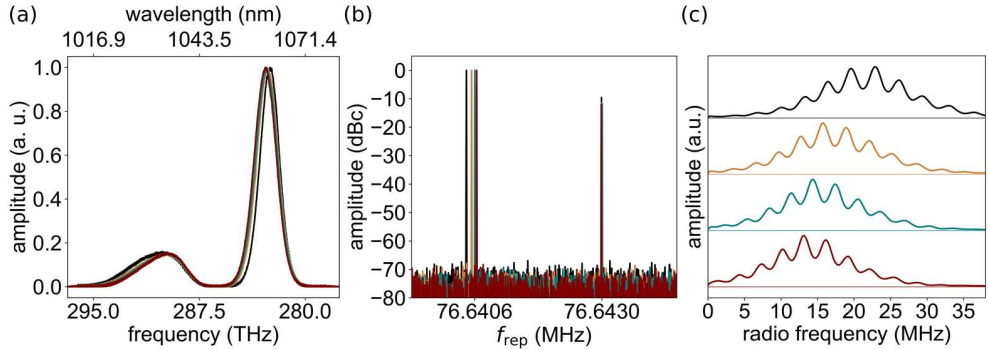

Fig. 4. (a) Spectral shift occurring when slightly changing the position of the beam block within the grating compressor using a micrometer screw. (b) Corresponding changes in the repetition rates due to the shift of the spectral filter. (c) Shift of the down-converted frequency comb along the radio-frequency axis.

### 4.1 Intra-cavity cross-talk

Dual-color fiber lasers are known to show amplitude fluctuations, which are generated by crosstalk caused by collisions of the two pulses inside the laser cavity [33]. However, the pulses inside the laser cavity recover and only a small temporal portion of the laser output is affected by this phenomenon.

Figure 5(a) shows the time domain output of our dual-comb system. In addition to the interferograms generated by optical beating between the two combs (Fig. 5(a) green section), spurious signals can be seen (Figure 5(a) red section). This spurious signal is present on both pulse trains individually and corresponds to a temporary modulation of the pulse amplitudes caused by the intra-cavity pulse collision. The position of the spurious signals relative to the interferogram depends on the extra-cavity path length difference between the 1030 nm and the 1060 nm-arms. If both arms have the same length, the interferogram and the spurious signals are overlapping. A path length difference of $\Delta l$ introduces a temporal shift $\Delta t_l$ between the two pulse trains, i.e.

$$\Delta t_l = \frac{\Delta l}{c}. \quad (4)$$

Hence, the position of the spurious signal from pulse train 1 will be shifted by $\Delta t_l$ on the temporal axis relative to the position of the spurious signal from pulse train 2. At the same time, the position of the center burst of the interferogram changes by

$$\Delta t_s = \frac{\Delta t_l}{\Delta t_c} \cdot \frac{1}{f_{rep,1}}, \quad (5)$$

where $\Delta t_c$ corresponds to the roundtrip time difference between the 1030 nm and the 1060 nm pulses, i.e.

$$\Delta t_c = |T_1 - T_2| = \left| \frac{1}{f_{rep,1}} - \frac{1}{f_{rep,2}} \right| = \left| \frac{\Delta f_{rep}}{f_{rep,1} f_{rep,2}} \right|. \quad (6)$$

For nominal repetition rates around 77 MHz, a $\Delta f_{rep}$ on the order of several kHz and a path length difference of some meters, $\Delta t_l$ amounts to less than 10 ns (and is thus not visible in Fig. 5), while $\Delta t_s$ on the other hand is on the order of a few hundred microseconds. Hence, by introducing an appropriate extra-cavity path length difference between the two pulse trains, we can maximize the non-disturbed time window $\Delta t_w$ around a center burst. This is important to maximize the resolution $\Delta v_{RF}$ of the current dual-comb setup, since the resolution $\Delta v_{RF}$ directly corresponds to the inverse of the measured time window $\Delta t_w$ as a consequence of the Fourier transform linking the time and frequency domain.

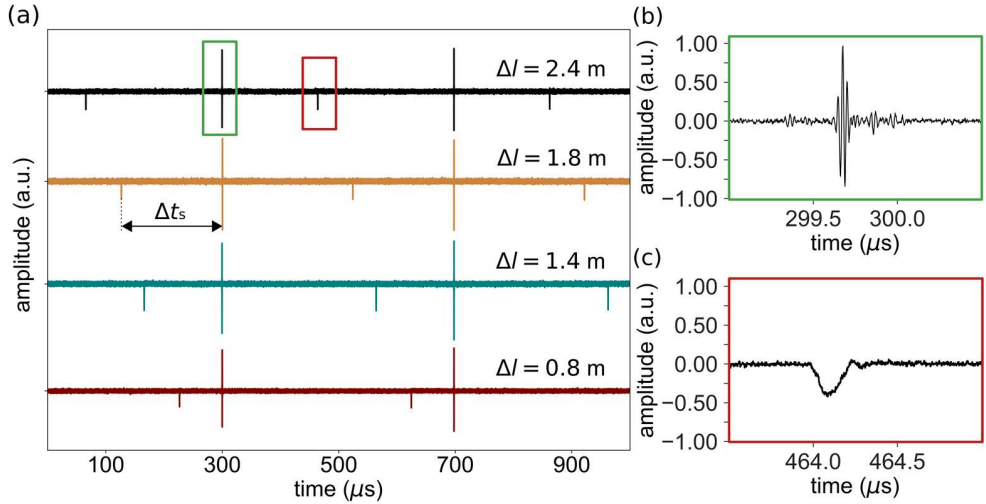

Fig. 5. (a) Time domain signal of the dual-comb recorded on an oscilloscope (LeCroy WavePro 760Zi). A linear interferogram can be seen (highlighted in green, zoom shown in (b)), as well as spurious signals caused by intra-cavity pulse collisions (highlighted in red, zoom shown in (c)). By changing the extra-cavity path length difference of the two arms (see setup in Fig. 1) the temporal position of the center burst can be shifted with respect to the position of the spurious signals.

## 4.2 Relative intensity noise (RIN)

Since amplitude noise can have a significant influence on the achievable signal-to-noise (SNR) in dual-comb measurements [34], we investigated the relative intensity noise (RIN) of the laser output at different positions in the setup: right after the spectral separation (before amplification and nonlinear broadening, see Fig. 6(a)), and right after amplification and nonlinear broadening (1030 nm pulse train) and the delay line (1060 nm pulse train), see Fig 6(b). The measurements were done by sending the beams onto a photodiode (PDA36-EC), low-pass filtering the output at 1.9 MHz and recording 100 time traces on a LeCroy WavePro 760Zi oscilloscope. The Fourier-transforms of the individual traces were averaged, corrected for the system noise floor and normalized by the DC-voltage output of the photodiode to yield the RIN spectra shown in Fig. 6. The amplitude noise caused by the intra-cavity pulse collisions is clearly visible in the RIN spectrum in the form of sharp peaks located at $\Delta f_{rep}$ = 2.6 kHz and its integer multiples. These side-peaks around -80 dBc are also visible when hooking up the photo diode output to a radio frequency analyzer (Keysight PXA N9030B) and zooming into the repetition rate signal, provided that the latter is displayed with an SNR > 80dB (Fig. 6(c)).

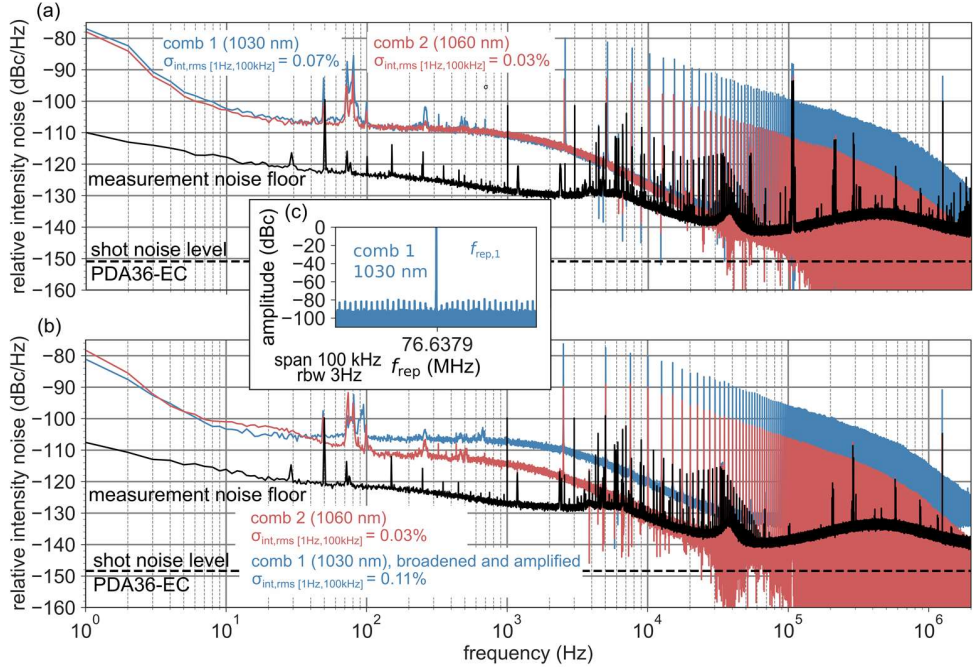

Fig. 6. (a) Relative intensity noise (RIN) of the spectrally separated combs measured before amplification including the root-means-quare (rms) RIN $\sigma_{int,rms}$ integrated over the interval [1 Hz, 100 kHz]. (b) RIN of the spectrally separated laser output after amplification and broadening including the root-means-quare (rms) RIN $\sigma_{int,rms}$ integrated over the interval [1 Hz, 100 kHz]. (c) Radio frequency spectrum of the broadened 1030-nm pulse train, showing a span of 100 kHz around the repetition rate signal with a resolution of 3 Hz.

There is a lack of RIN data in the literature for this kind of NALM Yb:lasers, even for standard single color/comb operation. However, we can compare our laser with well-documented state-of-the-art NPE mode-locked Yb:fiber lasers involving a free-space grating compressor for dispersion compensation [35]. For frequencies below 10 Hz, the RIN is about 10 to 20 dBc/Hz higher than reported in [35]. However, the noise at low frequencies can be significantly improved by increasing the stability of the free-space section by appropriate mechanical design. For frequencies above 10 Hz, the RIN is on the same order as for the other reported Yb:fiber lasers. The noise of the amplified and spectrally broadened light for low frequencies (mechanical vibrations) is the same as the low-frequency noise of the laser output, which is expected since the all-PM-fiber-amplifier design should not be influenced by mechanical vibrations. Above 100 Hz and especially for the cross-talk noise signals, a rise of the noise level due to amplification and nonlinear broadening is noticeable. This increase is on the same order of magnitude than described in previous amplifier systems [36].

### 4.3 Frequency stability

Another important property of the combs is their long-term frequency stability, in particular when the dual-comb system is meant to be used in a free-running state, i.e. without any active feedback. To characterize the long-term stability, we monitored the drift of the two repetition rates and their difference $\Delta f_{rep}$ over a duration of 200 min, see Fig. 7(a). Although a drift in the individual repetition rates is recognizable (standard deviation $\sigma(f_{rep,1}) = 12$ Hz and $\sigma(f_{rep,2}) = 13$ Hz), the drift of $\Delta f_{rep}$ is nearly one order of magnitude smaller (standard deviation of $\sigma(\Delta f_{rep}) = 1.7$ Hz, see Fig. 7(b)). This remaining drift could be further reduced by increasing the mechanical stability of the free-space portion of the laser, but was already small enough to allow for first proof-of-principle experiments as presented in the next section.

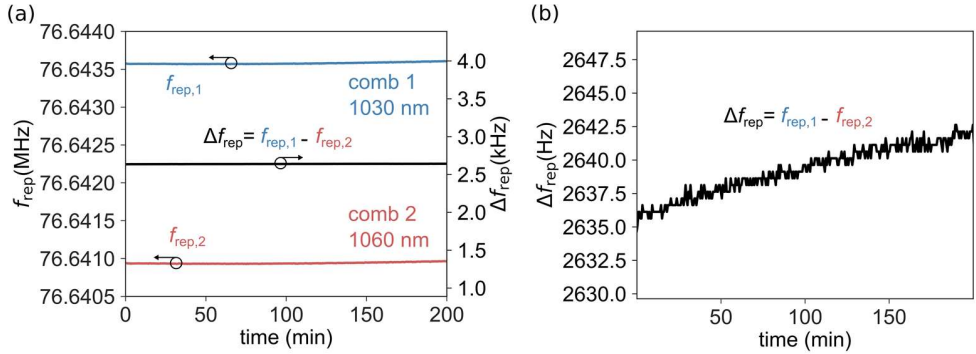

Fig. 7. (a) Drift of the individual repetition rate $f_{rep,1}$ and $f_{rep,2}$ as well as the difference $\Delta f_{rep}$ over 200 minutes, measured with a radio frequency analyzer (Keysight N9030B PXA). (b) Zoom into the drift of $\Delta f_{rep}$.

## 5. Etalon transmission measurement

To demonstrate the viability of spectral measurements, we measured the transmission of two etalons: an uncoated 700-μm thick gallium arsenide (GaAs) wafer and a 5-mm thick zinc selenide (ZnSe) window. For these measurements, 100 single bursts were recorded with a window size of 10 μs (GaAs) and 200 μs (ZnSe). These time traces were Fourier-transformed and averaged in the spectral domain directly using the built-in Fast-Fourier-Transform (FFT) and averaging functions of the LeCroy WavePro 760Zi oscilloscope (Fig. 8(a) and (e) orange curve). The data was then background-corrected and divided by a reference measurement (same measurement but without the etalon, Fig. 8(a) brown curve). In Fig. 8 (b), we show that the measured fringes are in good agreement with the theoretical transmission calculated for a 700-μm (GaAs) etalon (reflectivity $R = 0.30568$, refractive index $n = 3.4731$ at 1063 nm, [37]). As a sanity check and to obtain an absolute wavelength calibration, we also recorded the spectra in parallel with a calibrated grating-based optical spectrum analyzer (ANDO AQ6315A) that has a maximum resolution of 0.05 nm (Fig. 8(c)). Already in the case of the 700-μm thick GaAs sample, the resolution of the spectrum analyzer is insufficient to fully resolve the etalon transmission, which then leads to a slight discrepancy between the theoretical etalon curve and the transmission measured by the spectrum analyzer (Fig. 8(d)).

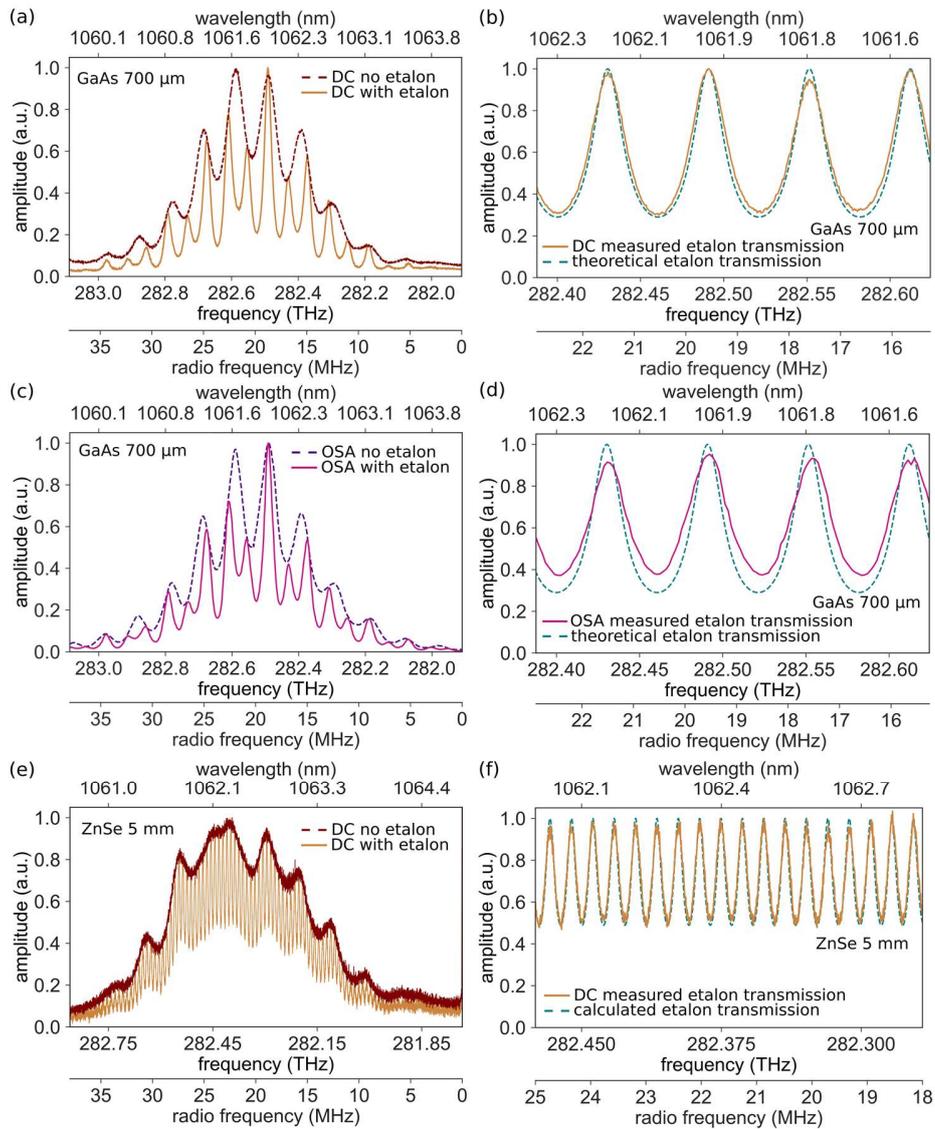

Fig. 8. Etalon transmission measurements. (a) Spectrum re-constructed from the dual-comb interferogram by averaging the FFTs of 100 single interferograms recorded in time windows of 10 μs each after a 3-nm band pass filter with and without a 700-μm GaAs etalon. (b) Measured transmission spectrum (orange, solid) obtained after background subtraction and division by the reference spectrum without the etalon, as well as theoretically calculated transmission function for a 700-μm GaAs wafer (turquoise, dashed). (c) Same measurement performed with an optical spectrum analyzer that has a maximum resolution of 0.05 nm (ANDO AQ6315A). This measurement was perfromed to obtain an absolute wavelength calibration for the spectra retrieved from the dual-comb data. Due to the limited resolution, the ANDO is not capable of fully resolving the fringes, as can be seen in (d). (e) Spectrum re-constructed from the dual-comb interferogram by averaging the FFTs of 100 single interferograms recorded in time windows of 200 μs each after a 3-nm band pass filter with and without a 5-mm ZnSe window. (d) Measured transmission spectrum (orange, solid) obtained after background subtraction and division by the reference spectrum without the etalon, as well as theoretically calculated transmission function for a 5-mm ZnSe window (turquoise, dashed).

In the case of the much thicker ZnSe etalon, the transmission fringes are beyond the resolution of the spectrum analyzer. However, the fringes measured with the dual-comb are again in good agreement with the theoretical transmission calculated for a 5-mm ZnSe etalon (reflectivity $R = 0.178$, refractive index $n = 2.458$ at 1063 nm, [38]) (Fig. 8(f)).The 200-µs window chosen to measure the ZnSe etalon implies a radio frequency resolution of 5 kHz. With a difference in repetition rate $\Delta f_{rep}$ of 2.6 kHz and a nominal repetition rate $f_{rep}$ of 77 MHz, this corresponds to an optical frequency resolution of 148 MHz. In order to obtain comb-line-limited resolution, the temporal window would need to be at least as long as $1/\Delta f_{rep} = 384$ µs. Since the spurious signals would be included in that case, some manipulation of the time window would be required to eliminate their influence. However, a manipulation such as zero-padding the spurious signals may obscure important spectral details in more complex spectroscopy measurements. Hence, a cleaner path towards achieving true comb-tooth resolution with this system would be to eliminate the spurious signals in a balanced-detection scheme, where half of the dual-comb light is sent through the sample, while the other half simultaneously provides a baseline-signal.

## 6. Conclusion and outlook

In summary, we have presented a novel method to obtain dual-color/dual-comb operation from a single all-PM Yb:mode-locked fiber laser. The method is based on mechanical spectral filtering and offers two main features: easy tuning of the difference in repetition rates (and hence of the tradeoff between non-aliasing bandwidth and measurement time), as well as the possibility to shift the down-converted frequency comb on the radio frequency axis by slightly tuning the position of the mechanical filter. Compared to previous all-PM single-cavity dual-color lasers based on erbium, we increased the nonaliasing bandwith by a factor of 2.5. Furthermore, we reached a factor of 5 higher output power than reported previously due to the high gain achievable in ytterbium gain fibers.

The combination of a fiber section exhibiting positive dispersion at 1 µm with a grating compression providing tunable negative GDD offers the possibility to make this laser run in different dispersion regimes. The impact of the dispersion regime on amplitude and frequency noise has been studied for NPE mode-locked fiber lasers [35], however no such study has been carried out yet for NALM lasers (let alone dual-color NALM lasers) to the best of our knowledge. Additionally, it is an open question whether and how the intra-cavity pulse collisions are influenced by the dispersion regime since pulse collisions have only been studied for dual-color solitons [33]. A detailed noise analysis for the different dispersion regimes of NALM lasers with nonreciprocal phase shifter in single- and dual-color mode has the potential to provide interesting insights. However, this is beyond the scope of this paper and will need to be addressed in future work.

The viability of our current setup was demonstrated in a proof-of-principle experiment by resolving the transmission fringes of two etalons (700-µm thick GaAs and 5-mm thick ZnSe) without the need for active stabilization of the comb parameters. In combination with recent work on computational averaging techniques [39], we believe that free-running single-cavity dual-comb approaches such as the one presented here are a promising path towards robust and compact high-precision spectrometers.

## Funding

Austrian Federal Ministry of Science, Research and Economy and the National Foundation for Research, Technology and Development.

Austrian Science Fund (FWF): M2561-N36

## Acknowledgment

The authors would like to acknowledge Garrett Cole and Dominic Bachmann for fruitful discussions and helpful comments on the manuscript, as well as Nikolai Kiesel for supporting us with measurement equipment.